\documentclass[12pt]{iopart}

\usepackage{graphicx}

\begin{document}
\title[Analytical relation between entropy production and
quantum Lyapunov...]{Analytical relation between entropy production
and quantum Lyapunov exponents for Gaussian bipartite systems}
\author{K. M. Fonseca Romero$^{2,3}$, J\'{u}lia E. Parreira $^1$, L. A. M. Souza$^1$, M. C. Nemes$^{1,3}$, and W. Wreszinski$^{3}$}
%\affiliation{$^1$ Address 1\\
%$^2$ address 2\\ $^3$ Address 3}
\address{$^{1}$Departamento de F\'{\i}sica, Instituto de Ci\^{e}ncias
Exatas, Universidade Federal de Minas Gerais, CP 702, CEP
30161-970, Belo Horizonte, Minas Gerais, Brazil;}
\address{$^{2}$Departamento de F\'{i}sica, Facultad de Ciencias,
Universidad Nacional, Ciudad Universitaria, Bogot\'{a}, Colombia;}
\address{$^{3}$Instituto de F\'{\i}sica, Universidade de S\~ao
Paulo, CP 66318 CEP 05389-970, S\~ao Paulo, SP, Brazil.}
\ead{carolina@fisica.ufmg.br}

\begin{abstract}
We study and compare the information loss of a large class of
Gaussian bipartite systems. It includes the usual Caldeira-Leggett
type model as well as Anosov models (parametric oscillators, the
inverted oscillator environment, etc), which exhibit instability,
one of the most important characteristics of chaotic systems. We
establish a rigorous connection between the quantum Lyapunov
exponents and coherence loss and show that in the case of unstable
environments, coherence loss is completely determined by the upper
quantum Lyapunov exponent, a behavior which is more universal than
that of the Caldeira-Leggett type model.
\end{abstract}


\submitto{\JPA}

\maketitle

\section{Introduction.} Application of Quantum Mechanics to real world many-body
systems meets  with several difficulties, both of conceptual and
pragmatic nature. A quantum mechanical system, which consists of at
least two interacting subsystems exhibits a completely nonclassical
property called ``entanglement''\cite{Schroedinger1935a}. It is a
nonclassical correlation between systems which exists even between
well separated subsystems \cite{Einstein1935a,Bell1964a}. This
unique property of a quantum system is nowadays viewed as a powerful
resource in quantum information theory and quantum computation
\cite{Nielsen2000a}. Also, on the conceptual side, ``entanglement''
with an environment (degrees of freedom which inevitably interact
with the quantum system of interest) is the key mechanism to explain
why typically quantum effects are not observed in macroscopic
systems \cite{vonNeumann1955a,Giulini1966a}. Therefore the rate at
which pure initial quantum states loose their potentiality to retain
information represents at the same time an old and a very modern
problem.

The first theoretical implementation of the system plus
environment dynamics, in both the weak and strong damping regimes,
was proposed and carried out by Caldeira and Leggett \cite{CaldeiraLeggett}.
They represent the environment as
an infinity of oscillators, weakly coupled to the system of
interest. Several successful descriptions of experiments,
specifically in quantum optics, were based on this type of model.
More recently questions have been raised about the role of
classical chaos on the information loss process for coupled
quantum systems. In particular, the role of chaos in the
decoherence process \cite{Haake2000a} has become a matter of
active research and also a matter of debate \cite{Zurek2001a,
Alicki2002a, Prosen2002a}.

One of the important issues
now is: which type of environment is the more effective: the one
involving an infinite number of degrees of freedom (quantum brownian
motion environment QBME), as proposed by Caldeira and Leggett, or an
environment consisting of a single degree of freedom which presents
one or more characteristics of chaotic behavior?

In the standard model of decoherence, a joint system consisting of a system (S)
 coupled to an environment (E) is described by a Hamiltonian
$$
H_J = H_S + H_E + V, \eqno{\rm{(1.1)}}
$$
where $H_S$ is the system's Hamiltonian, $H_E$ the environment's,
and $V$ the interaction Hamiltonian between S and E. Let, for
simplicity, the state space $\cal{H}_S$ of the system be
two-dimensional, and
$$
H_S = \lambda\sigma_{z}, \eqno{\rm{(1.2)}}
$$
with $\sigma_{z}$ the Pauli matrix with eigenvalues $+1$ and $-1$
corresponding to spin ``up'' and ``down'' along the z-axis, and
respective eigenfunctions $\psi_{+}$ and $\psi_{-}$. In the
coherent superposition
$$
\psi = \alpha_{+}\psi_{+} + \alpha_{-}\psi_{-}, \eqno{\rm{(1.3)}}
$$
with probability amplitudes $\alpha_{+}$ and $\alpha_{-}$ such
that $|\alpha_{+}|^2 + |\alpha_{-}|^2 = 1$, $\sigma_{z}$ has no
definite value. The environment state (Hilbert) space is
$\cal{H}_E$, and we suppose E is initially in the state
$\psi_E(0)\in \cal{H}_E$ and the joint system in the pure
(nonentangled) state
$$
\psi_J(0) = \psi \otimes \psi_E(0) \in \cal{H}_S\otimes\cal{H}_E.
\eqno{\rm{(1.4)}}
$$
We also assume for simplicity that the interaction $V$ in (1.1) is of the form
$$
V = \mu\sigma_{z}Q_E. \eqno{\rm{(1.5)}}
$$
Then, after a time $t$, the joint system will be in the state (units are chosen such that $\hbar=1$)
    $$
    \psi_J(t) = (\alpha_{+}\exp[-i\lambda
    t]\psi_{+}\otimes\psi_E^{+}(t) + \,
     \alpha_{-}\exp [i\lambda t]\psi_{-}\otimes\psi_E^{-}(t)),
    \eqno{\rm{(1.6a)}}
    $$
    with
    $$
    \psi_E^{\pm}(t) = \exp[-i(H_E\pm\mu Q_E)]\psi_E(0).
    \eqno{\rm{(1.6b)}}
    $$
We thus expect that the sensitivity to perturbation $\pm\mu Q_E$
of a chaotic environment Hamiltonian $H_E$ will, by (1.6a) and
(1.6b), result in quick decoherence:
$$
|(\psi_E^{+}(t), \psi_E^{-}(t))|\leq c\exp[-dt] \eqno{\rm{(1.7)}}
$$
where $d$ and $c$ are positive constants. The exponential
``orbital'' instability of $H_E$ is expected to lead via (1.6b) to
the exponential decay of the ``overlaps'' (1.7) (think of
Gaussians centered at the corresponding orbits): the decoherence
rate should therefore be proportional to the rate at which the
environment is able to explore its phase space. These
considerations are meant to explain, intuitively, why a (large)
collection of harmonic oscillators - being stable systems - could,
\textbf{in principle}, be less effective for purposes of
decoherence than one sole unstable system.

Recently, R. Blume-Kohout and W. H. Zurek \cite{Blume2003a}
analyzed decoherence due to a toy model for the environment, an
inverted harmonic oscillator environment (IHOE) -- see also
\cite{paz, arjendu1, arjendu2} where conclusions similar to ours
were obtained in chaotic systems. This is a nontrivial analytical
tractable model whose importance as realized in \cite{Blume2003a}
is that it shares one very important characteristic with chaotic
systems, namely exponential instability, defined by a positive
upper Lyapunov exponent \cite{Katok}. Systems of this sort are
called Anosov systems. We shall be rather dealing with generalized
Anosov systems, in the sense that the flows take place in a
manifold which is not compact. A simple example is the flow in the
$(p,q)$ plane defined (with $\lambda $ positive) by
$\frac{dp}{dt}=\lambda p$ and $\frac{dq}{dt}=-\lambda q$. The q
coordinates of any two points moving with the flow get closer and
closer as time proceeds, but the p coordinates separate
exponentially fast, and hence the two points move apart
exponentially fast \cite{Penrose}. The directions of stable and
unstable manifolds may vary from point to point, however, in
contrast to the above example: this is the case of the parametric
oscillator. The quantum analogues of the classical Anosov systems
have been dubbed \textbf{quantum Anosov systems} \cite{Emch}: they
exhibit a dynamic behavior which is at the same time rich and
universal and they are characterized by having a positive upper
quantum Lyapunov exponent, as defined in \cite{Jauslin} and in the
forthcoming (\ref{Lyapunov}), and will be the subject of the
present paper.
%( see Definition 6.4.2 of \cite{Katok} ).

Given the importance of the subject, the difficulty in obtaining
 mathematically sound exact results which are eventually able to shed
light onto these complex questions, we feel that a thorough
analytical investigation of the QBME vs. IHOE and its
generalizations to more realistic models of the class of linearly
coupled quantum Anosov systems, can be extremely useful. We focus
on the limited but important class of gaussian states. In this
context, a rigorous demonstration of the growth of the system's
reduced von Neumann entropy with the upper quantum Lyapunov
exponent for the aforementioned class of systems is given, showing
thus that the rate of information loss is of more universal nature
than that for the usual quantum brownian motion environment. Our
analytical results are only possible due to the fact that for
(possibly nonunitary) gaussian dynamics the rate of information
loss is completely governed by a combination of quadratures, i.e.
the covariance matrix, also called Schr\"odinger generalized
uncertainty principle.

\section{Relation between information loss and the covariance matrix
for gaussian states.} We start by recalling a well known result
\cite{agarwal,italianos}. The most general 1-D gaussian state can be
written as
\begin{equation}
\label{Gaussian}
  \hat{\rho}_{G}
  =
  \mathcal{D}(\alpha) \mathcal{S}(r,\phi)
  \hat{\rho}_{\nu}
  \mathcal{S}^\dag (r,\phi)
  \mathcal{D}^\dag (\alpha),
\end{equation} where \begin{eqnarray} \mathcal{D}(\alpha) &=& \exp \left( \alpha
\hat{a}^\dagger - \alpha^* \hat{a} \right), \\ \mathcal{S}(r,\phi)
&=& \exp \left( \frac{r}{2} \left( e^{i \phi} \hat{a}^{\dagger 2}
- e^{- i \phi} \hat{a}^2 \right) \right) \end{eqnarray} are the
displacement and the squeezing operator respectively and
$\hat{\rho}_{\nu}$ the thermal density operator with average
number of excitations $\nu$,
\begin{equation}
\label{rhonu}
  \hat{\rho}_{\nu}
   =
  \frac{1}{1+\nu}
  \exp\left(
    \ln\left(\frac{\nu}{\nu+1}\right)\hat{a}^\dag \hat{a}
  \right).
\end{equation}

Note that except in the improper limit of $\nu\rightarrow 0$,
which corresponds to a squeezed state, density matrix
\eref{Gaussian} is not a pure state. Its Schro\"odinger determinant
is  given by

\begin{eqnarray} \label{det}
D=\left(\begin{array}{ll}
\sigma_{pp}&\sigma_{qp}\\\sigma_{qp}&\sigma_{qq}
\end{array}\right)=(\nu+\frac{1}{2})^2
\end{eqnarray}
%\begin{align} \label{det}
%  D = \textrm{Det}
%\begin{pmatrix}
%    \sigma_{pp} & \sigma_{qp}\\
%    \sigma_{qp} &  \sigma_{qq}
%  \end{pmatrix}
%=(\nu+\frac{1}{2})^2
%\end{align}
where $\sigma_{xy}= \frac{1}{2} \tr \left({\hat
\rho}\left\{\hat{x},\hat{y}\right\}\right) -\tr(\hat{x}\hat
\rho)\tr(\hat{y}\hat \rho)$ and $\hat{x},\hat{y}$ are either the
position or its canonicaly conjugate momentum. We moreover have
for the (von Neumann) entropy
\begin{equation}\label{Entropia}
  S[\hat{\rho}]
   = -\tr\left(\hat{\rho}\ln \hat{\rho}\right) =
   (\nu+1)\ln(\nu+1)
   -\nu\ln \nu.
\end{equation}
Comparing \eref{Entropia} with \eref{det} we obtain the relation
between the entropy and the Schro\"odinger determinant
\begin{equation} \label{entdet}
S=\frac{1}{2}(\sqrt{D}+1)\ln(\sqrt{D}+1)
   -\frac{1}{2}(\sqrt{D}-1)\ln(\sqrt{D}-1))-\ln{2}.
\end{equation}

It is worthwhile to point out that \eref{entdet} holds for a wide
range of (gaussian) states, ranging from thermal (with
$\mathcal{S}(r, \phi)=1$, $\mathcal{D}(\alpha)=1$ in
\eref{Gaussian}) to generalized coherent states (the limit $\nu\to
0$ in \eref{rhonu}, with $\mathcal{S}=1$ in \eref{Gaussian});
\eref{entdet} is thus a fundamental relation for gaussian states.
%for the latter
%\begin{equation} \label{entline}
%\delta\equiv 1-\tr\left(\rho^2\right)=1-\frac{1}{\sqrt{D}}
%\end{equation}
%From \eref{entline}, $D=1$ corresponds to a pure state, which is
%consistent with the well known condition $S=0$ by \eref{entdet};
%again by \eref{entline}, $D>1$ corresponds to a mixed state.

\section{Quantum Brownian Motion Environment (QBME).} We next
consider an example of QBME widely used in the context of Quantum
Optics, whose dynamics (in the Born-Markov approximation) is
governed by the Liouvillian
\begin{eqnarray}
  \mathcal{L}
  &=&
  -i\omega \left[ \hat{a}^\dag \hat{a},\bullet \right]
  +k(\bar{n}_{B}+1)
  \left(
    2  \hat{a}\bullet \hat{a}^\dag
    -\hat{a}^\dag  \hat{a}\bullet
    -\bullet \hat{a}^\dag  \hat{a}\right) \nonumber \\ && +k\bar{n}_{B}
  \left(
    2 \hat{a}^\dag \bullet  \hat{a}
    - \hat{a}\hat{a}^\dag \bullet
    -\bullet  \hat{a}\hat{a}^\dag \right).
\end{eqnarray}
The entropy is a function of $\nu(t)$ only (see \eref{Entropia}),
where
\begin{equation}
\label{catorze}
 \nu(t) = \sqrt{(\sigma_{a\dag a}(t))^2-\sigma_{a\dag a\dag}(t)\sigma_{a a}(t)}-
\frac{1}{2},
\end{equation}
and
\begin{eqnarray}
 \sigma_{a\dag a}(t)
  &=&
  (\nu(0)+\frac{1}{2})\cosh(2 r(0)) e^{-2kt}
  +(\bar{n}_{B}+\frac{1}{2}) (1-e^{-2kt}),\\
 \sigma_{a\dag a\dag}(t)\sigma_{a a}(t) & =&  e^{-4kt}
  \left( (\nu(0)+\frac{1}{2})^2\sinh^2(2r(0))\right).
\end{eqnarray}

Notice that the entropy for this type of system may be obtained
without resorting to the calculation of the full density matrix,
due to the fact that the information exchange between the system
and bath is completely determined by a simple combination of
quadratures. Also the entropy saturates for long enough times.
Moreover, from equation \eref{catorze} it is apparent that
information loss depends both on the environment constants as well
as on the initial conditions.
%\begin{align}
%\label{IHE}
%\hat{H}
%= \sum_{i=1,2} \frac{1}{2} \hat{p}_i^2
%+\frac{\omega_1^2}{2} \hat{x}_i^2
%+\alpha\hat{x}_1\hat{x}_2,
%\end{align}
%\begin{align}
% \label{lambda}
%\pm\lambda=\pm&\sqrt{
%    \frac{1}{2} \left(
%      \sqrt{(1+\Lambda^2)^2+4\alpha^2}
%    +
%      1-\Lambda^2
%    \right)
%  },\\
%\label{omega}
%\pm\mu=\pm & i
%\sqrt{
%    \frac{1}{2} \left(
%      \sqrt{(1+\Lambda^2)^2+4\alpha^2}
%    -
%      1+\Lambda^2
%    \right)
%  },
%\lambda_{1,2} =
%\pm \sqrt{\frac{1}{2}}
%\end{align}

\section{Inverted Harmonic Environment (IHE).}
We now revisit the inverted harmonic oscillator example. Following (1.1),
we have now
\begin{eqnarray}
H_S &=& p_{1}^2/2 + \omega_{1}^2x_{1}^2/2 \\ H_E &=& p_{2}^2/2 +
\omega_{2}^2x_{2}^2/2 \\ V &=& \lambda x_{1}x_{2}
\end{eqnarray}
The above Hamiltonian  models two coupled
harmonic oscillators: the first one, is a regular oscillator, with
unit frequency, describing the system; the second one, an inverted oscillator with
squared frequency $\omega_2^2 = -\Lambda^2$, describing the environment. This is a nontrivial
tractable model. Its importance, as has been realized in reference
\cite{Blume2003a}, relies on its relation to chaos. Chaotic
behavior requires besides stretching and folding in phase space,
instabilities characterized by positive Lyapunov exponents.
Although the above hamiltonian is defined on the whole of phase
space and does not exhibit chaotic behavior, it displays positive
Lyapunov exponents like all Anosov systems. In this section we
establish a connection between Lyapunov exponents and the rate of
information loss, in this model, which we generalize to periodic
quadratic Anosov systems in the next section and exemplify with a
physically more sound system of coupled parametric oscillators.
The solution of the Heisenberg equations of motion for the vector
$\hat{\mathbf{z}} = (\hat{x}_1,\hat{p}_1,\hat{x}_2,\hat{p}_2)^T$
is as follows \cite{arthur}
\begin{equation}
\label{SolucaoIHE} \hat{\mathbf{z}}(t) = \mathbf{G} e^{\mathbf{L}
t} \mathbf{G}^{(-1)} \hat{\mathbf{z}},
\end{equation}
where $\mathbf{G}$ and $\mathbf{L}$ are constant matrices. The
former is invertible and the latter is diagonal with elements $
\pm\lambda=\pm A(\Lambda,\alpha)$ and $\pm \mu = \pm i
B(\Lambda,\alpha)$, where $A(\Lambda,\alpha)$ and
$B(\Lambda,\alpha)$ are \emph{real} and \emph{positive} constants.
The IHE Hamiltonian is a degenerate case of Floquet's theorem, see
the following. Notice the existence of an unstable mode, with
positive classical Lyapunov exponent $\lambda$. In reference
\cite{Jauslin}, upper quantum Lyapunov exponents were defined, for
systems of one degree of freedom with momentum $\hat{p}$ and
position $\hat{x}$, as follows
\begin{equation}
\label{Lyapunov} \bar{\lambda} = \sup_{\alpha\in\mathbf{R}^2}
\limsup_{t\rightarrow\infty} \frac{1}{t} \ln \Vert
[L_{\alpha},A(t,t_0)]\Vert
\end{equation}
were $\alpha=(\alpha_p,\alpha_x)$, and $A(t,t_0)$ is any bounded
operator, i.e.with finite norm given by $\Vert A \Vert =
\sup_{\psi} \Vert A\psi\Vert/\Vert\psi\Vert$, where $\psi$ belongs
to the Hilbert space $L_2(\mathbf{R}, \mathrm{d}x)$, and evolved
in the Heisenberg picture. This norm is nothing but the natural
extension of the usual matrix norm on the square integrable
sequences to an infinite-dimensional Hilbert space. The operator
$L_{\alpha}= \alpha_p \hat{p} +\alpha_x \hat{x}$ is the generator
of phase-space translations along a direction $\alpha$, and thus
\eref{Lyapunov} has a close resemblance to the definition of the
classical Lyapunov exponents (see, e.g., \cite{Alfredo1}),but
exploits the unitary nature of the quantum dynamics (see
\cite{Jauslin} for a discussion of these
points). %\sqrt{
%    \frac{1}{2} (
%      \sqrt{(1+\Lambda^2)^2+4\alpha^2}
%    +
%      1-\Lambda^2
%    )
%  },~
%\ \pm\mu=\pm  i \sqrt{
%    \frac{1}{2} (
%      \sqrt{(1+\Lambda^2)^2+4\alpha^2}
%    -
%      1+\Lambda^2
%    )
%  }

Also in reference \cite{Jauslin} the definition \eref{Lyapunov}
was applied for one of the simplest paradigms of the transition
from regular to unstable behavior in classical mechanics, namely
the parametrically driven oscillator, given by
\begin{equation}\label{HFt}
 H(t) = \frac{1}{2} \hat{p}^2 +\frac{1}{2} f(t) \hat{x}^2,
\end{equation}
where $f(t)$ is a periodic function. The dynamical solution  for
$\hat{x}(t,0)$ and $\hat{p}(t,0)$ is a formally one dimensional
version of \eref{SolucaoIHE}, but with $\mathbf{G}$ being a
periodic matrix with the same period of the driving and
$\mathbf{L}$ a traceless matrix: this is Floquet's theorem and the
eigenvalues of $\mathbf{L}$ are known as Floquet exponents (see
\cite{Jauslin} for the proof and references). Applying definition
\eref{Lyapunov} it was found that the upper quantum Lyapunov
exponent is precisely the real part of the maximal Floquet
exponent, which is positive in the classical instability region of
the parametric oscillator, and zero in the stability region. The
generalization to two degrees of freedom is straightforward
\cite{Sapin} and, applied to the IHE shows that the upper quantum
Lyapunov exponent \eref{Lyapunov} is $\lambda$ (see equation
eingenvalues of the IHE). Let now $S_{r}= -\tr\left(
\rho_{1}\ln{\rho_{1}}\right)$, the von Neumann entropy of the
reduced density matrix, with $\rho_{1}\equiv tr_{2}\rho$, with
$\rho$ of the form \eref{Gaussian}-\eref{rhonu}, but now with $\nu
\to 0$, i.e., the initial state being a tensor product of
generalized coherent states. Since $\rho_{1}$ is also Gaussian, it
continues to satisfy \eref{entdet}. In both cases, IHE and
parametric oscillator \eref{HFt}, the determinant is found to be
of the form
\begin{equation}
\label{determinant}
D(t) = \sum_{a,b=-2}^2 C_{ab} e^{(a\lambda+b\mu)t},
\end{equation}
where the constant coefficients $C_{ab}$ are given by cumbersome
not particularly enlightening expressions. These coefficients,
which are zero unless the sum of $a$ and $b$ is an even number,
depend on the variances of the initial state and will be omitted
here. For instance, one can find $C_{a b}$ by looking at the
asymptotic behavior of \eref{Entropia}. For large enough times $t$
we have $D(t)\sim C_{20} e^{2\lambda t}$ and the reduced entropy
can be approximated by $S_{r}\sim \ln(\sqrt{D}-1/2)+1\sim
\ln{D}/2$. This behavior,
\begin{equation} \label{entparam}
 S_{r}(t)\sim \ln(C_{20})/2 +\lambda t
\end{equation}
is precisely the one conjectured by Zurek several years ago
\cite{Blume2003a}. Since $\rho_{1}$ is also gaussian,
\eref{entdet} applies to $\rho_{1}$ and thus the linear growth of
$S_{r}(t)$ is interpreted, in the present case of a bipartite
system, as a coherence loss, which, as we see, \textbf{depends
only on the upper quantum Lyapunov exponent} (uqLe) and is thus
universal for this class of systems, in contrast to QBME. Note
that complexity in the present models (they are not explicitly
soluble classically and are a paradigm of the transition from
regular to irregular behavior, see \cite{arnold}) is brought about
by the external field, which depends on time in a nonlinear way!
For the models with frequency varying almost periodically with
time - included in the present treatment - see \cite{29}: it may
be considered as a prototype of complexity. This is one of the
reasons why our extension of \cite{Blume2003a} is significant:
Zurek's IHOE is \textbf{not} complex!

As remarked in \cite{Blume2003a}, section II, p. 032104-2, the IHE
Hamiltonian is unphysical in that in general the directions of
stable and unstable manifolds can vary from point to point. Thus
it is of special interest to be able to consider as system 2 an
arbitrary quantum Anosov system. In what follows we illustrate
this by taking both systems of the form \eref{HFt}, a case in
which most computations can be done explicitly. This model is
particularly rich: there is even a transition from a stability to
an instability region, which has a physical interpretation in a
model of quadrupole radio-frequency traps (Paul-Penning traps)
(see \cite{Jauslin} and references given there).
%\begin{equation}
%S_{r}=-\tr\left(\rho_{1}\ln{\rho_{1}}\right)
%\end{equation}
%\begin{align} \label{Coupled} \notag
% H(t)
%= \sum_{i=1,2} \frac{1}{2} \left( \hat{p}_i^2 + \left(\omega_i^2 -
%q \cos 2t \right)\hat{x}_i^2\right) +\frac{g}{2}
%(\hat{x}_1-\hat{x}_2)^2.
%\end{align}
%\begin{align}
% u_1(t) \hat{x}_1(0) +u_2(t) \hat{p}_1(0) +u_3(t) \hat{x}_2(0) +u_4(t) \hat{p}_2(0),
%\end{align}
%\begin{align}
% C(\alpha,q,t) =\frac{1}{2}\left(
%e^{i\phi t} P(\alpha,q,t)+e^{-i\phi t} P(\alpha,q,-t)
% \right),
%\end{align}

For the sake of comparison, in figure 1 we plot the entropy
obtained for a quantum Brownian motion environment (BME) and for a
IHO reservoir (parameters: (i) for the BME $\nu_0 = 1$, $r_0=1$,
$k=0.5$, $n_B=10$, $\omega=1$, $\phi=0$; (ii) for the IHO
$\ln(C_{20})/2=1.4$ and $\lambda = 1$). Note that the increasing
in the entropy in the case of BME saturates when the system
thermalizes with the reservoir, while the increase in the case of
IHO is linear (in the asymptotic regime).

\begin{figure}
\begin{center}
\includegraphics[width=8cm, height =6cm ,angle=0]{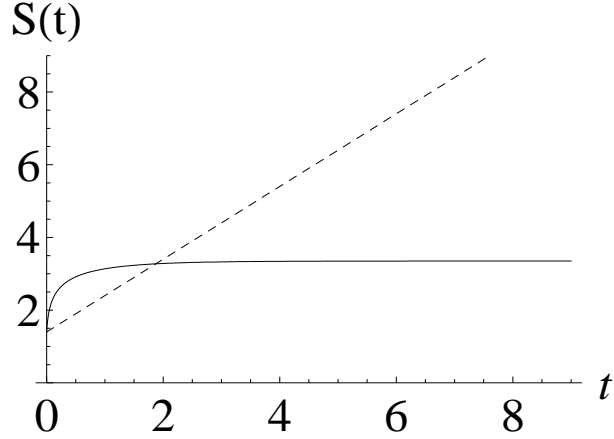}%72 -26 541 719
\end{center}
\caption{Time evolution of the entropy for a quantum Brownian
motion environment (solid curve) and for a inverted Harmonic
Oscillator reservoir (dash curve).}
\end{figure}

\section{Coupled parametric oscillators.} Consider now a more
realistic example, physically relevant in connection with Paul
traps \cite{Casati}, which models two coupled parametric
oscillators with hamiltonian \begin{equation} \label{htcoupled}
H(t) = H_S(t)+H_E(t)+V \end{equation} where \begin{eqnarray} H_S
(t) &=& \frac{1}{2} \hat{p}_1^2 + \left( \omega_1^2 - q \cos 2 t
\right) \hat{x}_1^2, \\ H_E(t) &=& \frac{1}{2} \hat{p}_2^2 +
\left( \omega_2^2 - q \cos 2 t \right) \hat{x}_2^2 \end{eqnarray}
and
\begin{equation} V = \frac{g}{2} (\hat{x}_1-\hat{x}_2)^2.
\end{equation} In what follows we assume $g>0$ and
$\omega_2^2>\omega_1^2$. We remark that \eref{htcoupled}
corresponds to the standard model (1.1). The solution of the
Heisenberg equations of motion for the position and momentum
operators can be written in terms of Mathieu functions. For
example, the momentum of the second oscillator, $\hat{p}_2(t)$ is
given by the expression $ u_1(t) \hat{x}_1(0) +u_2(t) \hat{p}_1(0)
+u_3(t) \hat{x}_2(0) +u_4(t) \hat{p}_2(0), $ where the functions
$u_a(t),~ a=1,2,3,4$ are given, respectively, by
\begin{eqnarray}
u_1&=& \sin(2\theta)\left[ \frac{\dot{S}(\alpha_-,q,0)}{2 D_2}
C(\alpha_-,q,t) -\frac{\dot{S}(\alpha_+,q,0)}{2 D_1}
C(\alpha_+,q,t)
 \right] \\ u_2&=&
\cos^2\theta\frac{\dot{S}(\alpha_-,q,0)}{D_2} C(\alpha_-,q,t)
-\sin^2\theta\frac{\dot{S}(\alpha_+,q,0)}{D_1} C(\alpha_+,q,t)
\\ u_3&=&
\sin(2\theta)\left[ \frac{C(\alpha_-,q,0)}{2D_2} S(\alpha_-,q,t)
-\frac{C(\alpha_+,q,0)}{2D_1} S(\alpha_+,q,t)
 \right] \\ u_4&=&
\cos^2\theta\frac{C(\alpha_-,q,0)}{D_2} S(\alpha_-,q,t)
-\sin^2\theta\frac{C(\alpha_+,q,0)}{D_1} S(\alpha_+,q,t).
\end{eqnarray}
The parameters  $\alpha_{\pm} =
\frac{\omega_1^2+\omega_2^2}{2}+g\pm
\sqrt{g^2+\frac{(\omega_2^2-\omega_1^2)^2}{4}}$ depend on the
constants appearing in the Hamiltonian. The angle $\theta$
appearing in the expressions above is determined from the
equality: $\tan(2\theta)=\frac{2g}{\omega_2^2-\omega_1^2}$. The
functions $C(\alpha,q,t)$ and $S(\alpha,q,t)$ are the usual
Mathieu  cosine and sine functions, and $\dot{C}(\alpha,q,t)$ and
$\dot{S}(\alpha,q,t)$ their time derivatives. From the expression
$ C(\alpha,q,t) =\frac{1}{2}\left( e^{i\phi t}
P(\alpha,q,t)+e^{-i\phi t} P(\alpha,q,-t)
 \right),
$ where $P(\alpha,q,t)$ is a periodic function with period $\pi$
and $\phi=\phi(\alpha,q)$ is the so-called characteristic
exponent, it can be seen that, if $\Im(\phi)\neq 0$ the solutions
exhibit an unstable behavior. Otherwise the solutions are
uniformly bounded. We are interested here in a situation in which
the second uncoupled forced oscillator (system) is in the
stability region and the other (environment) is in the unstable
region. From the solutions of the Heisenberg equations of motion
and the definition of the upper quantum Lyapunov exponent
$\bar{\lambda}$, it is a simple matter to show that $\bar{\lambda}
= |\Im(\phi(\alpha_1,q))|$. The connection with the information
loss is made through the evaluation of Schro\"odinger determinant
for the second oscillator, which turns out to be of the form of
equation \eref{determinant}. In this case, however, the
coefficients $C_{ab}$ are periodic functions of time. Moreover
$\bar{\lambda}=|\Im(\phi(\alpha_1,q))|$ and $\mu=i
\phi(\alpha_2,q)$. For an initial gaussian state,  $S(t_n)\sim
\ln(C_{20})/2 +\lambda t_n$, where $t_n=2\pi n$. The coefficients
$C_{a b}$ are obtained from the asymptotic behavior of the entropy
\eref{Entropia}.

\section{Conclusions.} In spite of exhibiting a complex, rich
dynamical behavior, bipartite open quantum Anosov systems display,
in their instability region of parameters, a reduced von Neumann
entropy with linear growth determined by their uqLe, a universal
behavior for this class of systems, in contrast to QBME. This also
permitted us to find a unified description of two opposite
regimes, viz. of small and large coupling. It may, however, be
argued that \eref{determinant} is \textbf{induced} by classical
mechanics: the value of the uqLe for the present models equals the
classical maximal Floquet exponent, which is positive in the
region where the classical parameter $E=\frac{1}{T}\int_0^T
f(t)\,dt$ lies in the instability region (``gap'') of Hill's
equation $\frac{d^2 x}{dt^2}+f(t)x=0$ \cite{Hochstadt}, as may
abstracted from the last section. This is a consequence of the
fact that the dynamics of the \textbf{observables} is classical.
The space of \textbf{states} is, nevertheless, subject to the laws
of quantum mechanics, without classical analogue, even in these
cases: for this reason, for gaussian initial states, the entropy
growth \eref{entparam} is to be interpreted, due to \eref{entdet},
as coherence loss, a purely quantum mechanical concept. Indeed, in
contrast to quantum mechanical entropy, the classical entropy is
not an adequate quantity in information theory, because it is not
even \textbf{positive} in general: (\cite{37}, Prop.1). % it tends, e.g.,
%to $-\infty$ as $T\to 0$ for the classical ideal gas (see any
%standard text in thermodynamics).
Negative entropies are due to the fact that classical density
distributions may be concentrated in regions of phase space
$<h$,contradicting the uncertainty principle (\cite{34},
\cite{35}). Modified ``semiclassical entropies'' without this
inconvenience have been defined by Wehrl \cite{35}, but they
incorporate the Hilbert space structure of quantum physics, i.e,
of a countable discrete basis of states. Entanglement has indeed
been described in terms of this modified entropy \cite{36}. Thus
an explanation by ``classical entanglement'' of phenomena of
information loss is not possible in a full classical treatment
involving the entropy.

A still different approach is to regard entanglement in terms of
the Wigner function, i.e., in phase space, which illuminates quite
different and surprising aspects \cite{alfredo2}.

We should caution the reader that, even in the framework of
(time-dependent) quadratic Hamiltonians, our approach misses an
important element which is present in fully chaotic systems:
\textbf{folding}, which is characteristic of a compact phase
space. Indeed, other papers have reported on the growth of
relative entropy in the context of quantized chaotic systems
(\cite{31},\cite{32}). In contrast to the present treatment they
rely, however, on approximations (based on random matrix theory in
\cite{31}, semiclassical analysis in \cite{32}), numerical studies
and certain special constructions (the nonunitary step in
\cite{32} is constructed to mimic the presence of diffusion) and
special concepts (a particularly phase-space measure of complexity
was used in \cite{31}). We hope that the discrete Weyl-Wigner
formalism developed in \cite{galet} may be used to extend our
treatment to some quantum chaotic systems.

As a final remark, we mention that other parameters could be invoked
for environments characterized by generic chaotic systems: the
diffusion coefficient is one of them, which is conjectured to be
related to the Lyapunov exponent and the Hausdorff dimension
\cite{33}.

\ack Parreira, Souza  Nemes and Wreszinski thank CNPq-Brasil for
financial support. We should also like to thank the referees for
illuminating remarks and suggestions which considerably improved
the understanding of the text.

\end{document}